\documentclass[12pt]{iopart}

\usepackage{iopams}  
\usepackage{graphicx}

\newcommand{\Aslash}{A \hskip -0.6em /}
\newcommand{\dslash}{\partial \hskip -0.6em /}
\newcommand{\Det}{\mbox{Det}}
\begin{document}

\title[Energies of Quantum QED Flux Tubes]
{Energies of Quantum QED Flux Tubes\footnote[7]{\hspace{-0.1cm}Talk 
presented at the QFEXT 05 workshop in Barcelona, Sept. 2005. To appear
in the proceedings.}}

\author{H Weigel}

\address{Fachbereich Physik, Siegen University,
Walter Flex Stra{\ss}e 3, 57072 Siegen, Germany}
\ead{weigel@physik.uni-siegen.de}
\begin{abstract}
In this talk I present recent studies on vacuum polarization 
energies and energy densities induced by QED flux tubes.
I focus on comparing three and four dimensional scenarios 
and the discussion of various approximation schemes in view
of the exact treatment.
\end{abstract}


\section{Introduction and Motivation}

In this talk I will present calculations of vacuum polarization
energies that we~\cite{Gr04} have performed for flux tube 
configurations in QED.
Flux tubes in QED coupled to fermions exhibit interesting
phenomena, such as the Aharonov-Bohm effect \cite{Aharonov:1979}, its
consequences for fermion scattering \cite{Alford89}, parity
anomalies~\cite{Redlich:1984dv}, formation of a 
condensate~\cite{Ca95}, and exotic quantum 
numbers~\cite{Blankenbecler:1986ft,Kiskis97,Niemi:1983rq}.  
Those (non-perturbative) features of the theory
that give rise to these unusual phenomena make the analysis 
more difficult, especially in calculations
that require renormalization. The investigation in
ref.~\cite{Bordag:1998tg} and the world line formalism in
ref.~\cite{Langfeld:2002vy} have addressed some of these issues. Here we
provide a comprehensive approach employing techniques from scattering
theory to analyze quantum energies of flux tubes.

Our primary motivation for this analysis is to shed light on
vortices in more complicated field theories, especially the Z--string
in the standard electroweak theory~\cite{Achucarro:1999it}. 
The Z--string is a vortex configuration carrying magnetic flux in 
the field of the Z-gauge boson.  Since the classical 
Z--string is known to be unstable \cite{Vachaspati:1992fi}, 
the existence of such a vortex
would require stabilization via quantum effects~\cite{Farhi:2000ws}, 
perhaps by trapping heavy quarks along the string. 

We compare the one-loop energies and energy densities of electromagnetic
flux tubes in $D=2+1$ and $D=3+1$ spacetime dimensions. The classical
calculation is of course the same in the two cases. The quantum
corrections to the energy could possibly be very different
\cite{Langfeld:2002vy} because of the different divergence structure.
In $D=3+1$, the bare one-loop energy is divergent and requires
renormalization. In $D=2+1$, in contrast, the bare energy is finite.  
However, a comparison between the two dimensionalities is sensible only 
when we use the same renormalization conditions, which induces a finite 
counterterm in the $D=2+1$ case. Without this finite renormalization, 
the $D=2+1$ and $D=3+1$ energies are qualitatively different.  

We also study this problem to analyze several technical puzzles 
associated with the computation of the one--loop energy of a flux tube.  
An efficient way to compute the energy is to use scattering data of
fermions in the background of the flux tube.  However, vortex
configurations give long--range potentials, that do not satisfy standard
conditions in scattering theory \cite{Newton:1982qc}, which usually
guarantee the analytic properties of scattering data. In turn these
properties are crucial to compute the vacuum polarization energy
from scattering data. Hence we observe
subtleties that emerge only because an isolated flux tube is
unphysical, and once a region of return flux is included, the
scattering  problem is well-defined. In the
limit where the return flux is infinitely spread out, the energy 
density becomes entirely localized at the original flux tube.

\section{Theory}

We consider the QED Lagrangian
\begin{equation}
\mathcal{L}=-\frac{1}{4}\left(\partial_\mu A_\nu-\partial_\nu A_\mu\right)^2 
+\bar{\psi}(i\dslash-e\Aslash-m) \psi\,,
\label{qedlag}
\end{equation}
where $A_\mu$ is the Lorentz vector that represents the photon
field and $\psi$ the spinor of a fermion which electric charge $e$. 
In $D=3+1$ $\psi$ is required to have four components. Though that
is not necessary in $D=2+1$ we may nevertheless choose so.

In radial gauge the flux tube configuration reads:
\begin{equation}
A_0=0\,, \qquad \vec{A}=\frac{F}{2\pi r}f(r)\hat{e}_{\varphi}
\,, \qquad
B (r) =\frac{F}{2\pi r}\frac{df(r)}{dr}\,,
\label{radgauge}
\end{equation}
where $B$ denotes the magnetic field.
For a Gau{\ss}ian flux tube we thus find
\begin{equation}
f_G(r)=1-e^{-r^2/w^2} \,, \qquad
B_G(r)=B_G(0) e^{-r^2/w^2} \,, 
\label{gaussft}
\end{equation}
which implies the net flux $F_G=\pi w^2B_G(0)$ and the classical
energy,
\begin{equation}
E_{\rm cl} = \frac{1}{2} \int d^2 r B^2(\vec{r\,})
\label{ecl1}
\end{equation}
in $D=2+1$. For $D=3+1$ the identical expression gives the energy 
per unit length of the flux tube. We adopt the one loop approximation 
wherein the background fields consists only of gauge fields; 
so do the external 
lines of the Feynman diagrams. Thus we only consider fermions loops in 
the one loop approximation and the only relevant counterterm Lagrangian is
\begin{equation}
{\cal L}_{\rm ct}=-\frac{C^{(D)}}{4} F_{\mu\nu}F^{\mu\nu}\,.
\label{lct}
\end{equation}
It is natural to impose the renormalization condition that
at zero momentum transfer the photon wave--function is not
altered by quantum effects. This yields
$C^{(3)}=-\frac{e^2}{6\pi m}$ and $C^{(4-\epsilon)}=-\frac{e^2}{12\pi^2}
\left(\frac{2}{\epsilon}-\gamma+\ln\frac{4\pi}{m^2}\right)$ in
$D=2+1$ and $D=3+1-\epsilon$ dimensions respectively. Note that
the renormalization coefficient is finite in $D=2+1$ while we 
have employed dimensional regularization in $D=3+1$.
The corresponding counterterm energy (resp. energy per unit length) is
\begin{equation}
E^{(D)}_{\rm ct}=\frac{C^{(D)}}{2}\,\int d^2 x B^2\,.
\label{ect1}
\end{equation}

\section{Fermion Determinant}

We obtain the vacuum polarization energy from the fermion loop
by computing the functional determinants
\begin{eqnarray}
E_{\rm vac}^{(3)} &=& \lim_{T \rightarrow \infty} \frac{i}{T}
\left[ \ln \Det (i\dslash - e\Aslash - m) - \ln \Det (i\dslash - m)
  \right] + E^{(3)}_{\rm ct} \cr
E_{\rm vac}^{(4)} &=& \lim_{T,L_z \rightarrow \infty} \frac{i}{TL_z}
\left[ \ln \Det (i\dslash - e\Aslash - m) - \ln \Det (i\dslash - m)
  \right] + E^{(4)}_{\rm ct}\,.
\label{fermiondet}
\end{eqnarray}
Again, we consider the energy per unit length of the flux tube. Note
that by inclusion of the counterterm contribution the above 
expressions are ultra--violet finite. In order to perform this
computation we have to consider the non--trivial (static) background 
field $A_\mu(\vec{x\,})\ne0$ of the flux tube, 
{\it cf.\@} eqs.~(\ref{gaussft}) in the Dirac equation:
\begin{equation}
\left[\vec{\alpha\,}\cdot\left(\vec{p\,}+\vec{A\,}(\vec{x\,})\right)
+\beta m\right]\Psi =\omega\Psi\,.
\label{Diraceq}
\end{equation}
The interaction induces a potential for the fluctuating fermions 
fields with two essential properties.  First, bound states with 
energies $\omega_j$ may emerge. In magnitude these energies are smaller 
than the fermion mass $m$\footnote[7]{For the flux tube configurations
only threshold states with $|\omega_j|=m$ occur. The configurations
that we consider in section 4 does not have any bound states.}. 
Second, the continuum levels 
$\omega=\pm\sqrt{k^2+m^2}$ acquire a non--zero phase shift, 
$\delta_\ell(k)$ which translates into a modification of the level 
density, $\Delta \rho(k) = \frac{1}{\pi} \sum_{\ell,\pm} 
\frac{d}{dk}\left[\delta_\ell (k)\right]$. Here $\ell$ is the
orbital angular momentum quantum number according to which the 
modes decouple and $k$ is the linear momentum of the fluctuating
field. We then find the vacuum polarization energy

\parbox[t]{12cm}{\hskip-1.5cm
\begin{minipage}[l]{9cm}
\begin{eqnarray}
E^{(3)}_\delta & = &
-\frac{1}{2} \sum_j\left(|\omega_j|-m\right)
+ \frac{1}{2\pi} \int_0^\infty dk\,
\frac{k}{\sqrt{k^2+m^2}} \sum_\ell \bar{\delta}_\ell(k) 
\nonumber \\
E^{(4)}_\delta & = &
-\frac{1}{8\pi}\sum_j\Big(\omega_j^2\ln \frac{\omega_j^2}{m^2}
+m^2-\omega_j^2\Big)
- \frac{1}{4\pi^2} \int_0^\infty dk k \ln
\frac{k^2+m^2}{m^2} \sum_\ell \bar{\delta}_\ell (k) 
\nonumber \\
\bar{\delta}_\ell(k)&=&\delta_\ell (k) - \delta_\ell^{(1)} (k) -
\delta_\ell^{(2)}(k)\,.
\label{vacpoleng}
\end{eqnarray}
\end{minipage}}

\smallskip \smallskip \noindent
In $D=3+1$ the energy per unit length is obtained from the 
interface formalism of ref.~\cite{Gr01}. It is important to stress
that we have subtracted the first two orders of the Born series
from the integrand to render the integrals finite. We will add back 
in these pieces in from of Feynman diagrams, $E^{(D)}_{\rm FD}$. 
The identity between Born and the Feynman diagram contributions
at a prescribed order has been verified within dimensional 
regularization~\cite{Fa00} and tested numerically, see {\it e.g.}
appendix B of ref.~\cite{Fa01}. This
procedure has the advantage that the renormalization conditions
from the perturbative sector of the theory may be adopted~\cite{Gr02}.
To this end the renormalized vacuum polarization energy reads
\begin{equation}
E^{(D)}_{\rm vac}=E^{(D)}_\delta+E^{(D)}_{\rm FD}+E^{(D)}_{\rm ct}\,.
\label{evacren}
\end{equation}

\section{Embedding}

\begin{figure}[b]
\begin{center}
\includegraphics[width=6.0cm,height=4.5cm]{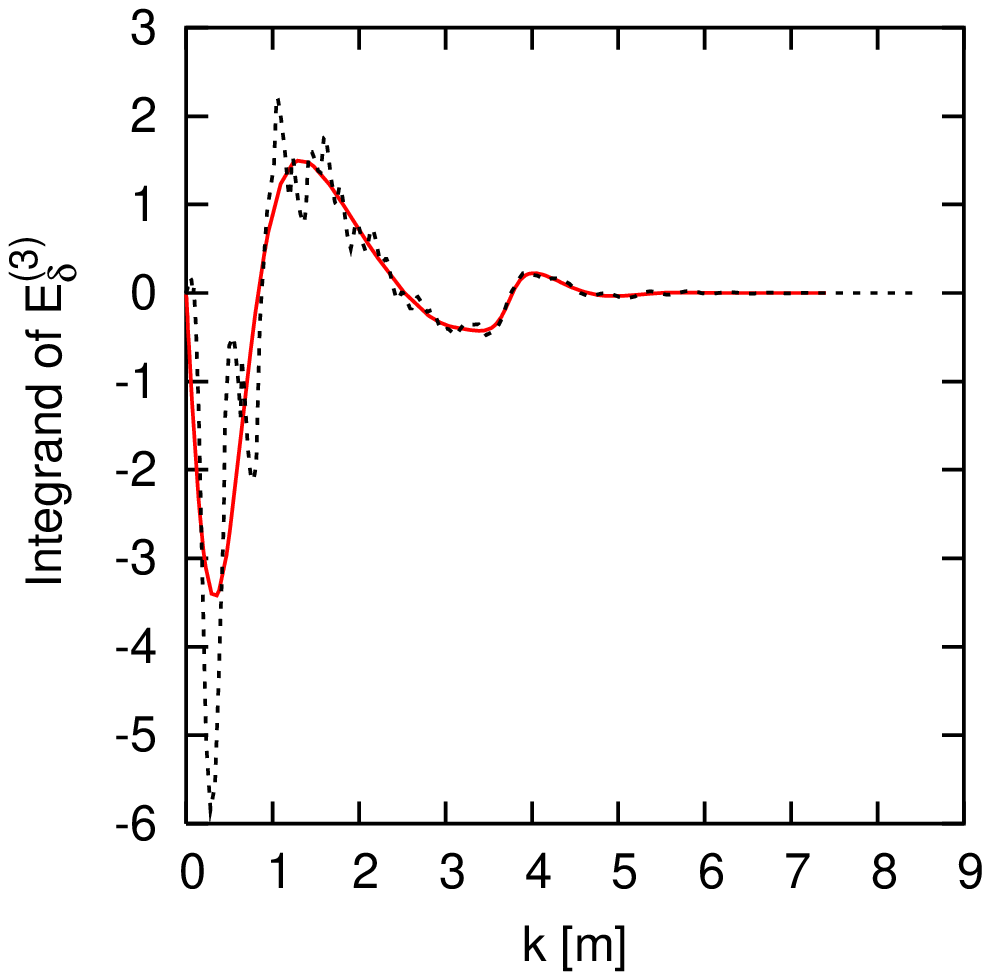}
\hskip 1.0cm
\includegraphics[width=6.0cm,height=4.5cm]{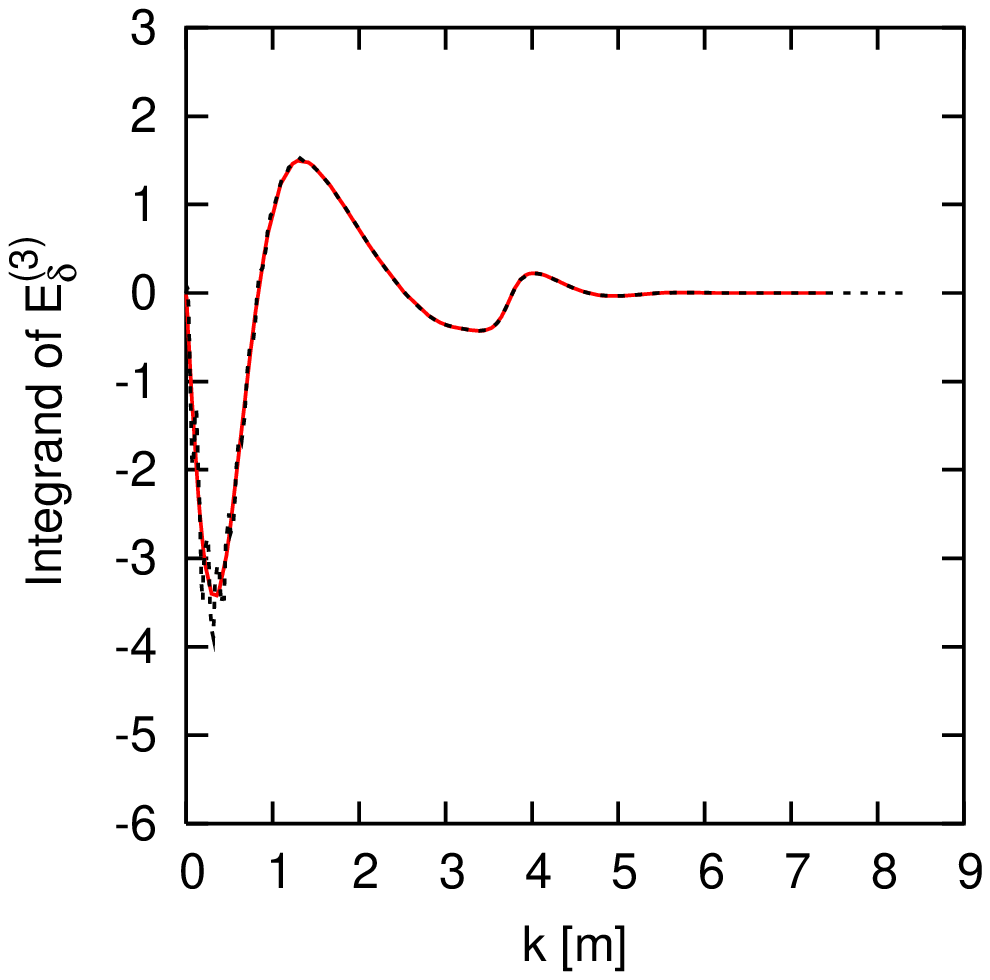}
\caption{\label{fig_1}The integrands of $E_\delta^{(3)}$,
eq.~(\ref{vacpoleng}) with (black dashed lines) and without (red
full lines) return fluxes for $R=6/m$ (left panel) and $R=26/m$
(right panel).}
\end{center}
\end{figure}

Configurations like eq.~(\ref{gaussft}) with non--zero net flux
induce potentials that behave like $V_{\rm eff}(r)\sim1/r^2$
as $r\to\infty$ in the second order differential equations 
for the radial functions from which we extract the bound state 
energies and phase shifts. This behavior violates standard conditions 
necessary to deduce the analytic structure of scattering data 
in the complex momentum plane. As a direct consequence we observe 
that the phase shifts are discontinuous as $k\to0$ and Levinson's 
theorem cannot be employed to reliably predict the number of bound 
states. Though this is not a principle obstacle because we have 
other means to find the bound states and any singularity at $k=0$ is 
integrable in eq.~(\ref{vacpoleng}) it puts doubts on the use of 
scattering data for this computation. The analytic structure is 
furthermore mandatory to relate the matrix element of the energy 
momentum tensor to formulas like eqs.~(\ref{vacpoleng}) and~(\ref{evacren})
that underly our computation of the renormalized vacuum polarization
energy.  At the same time we observe that configurations with 
zero net flux are \emph{unrealistic}. This becomes obvious from the 
Bianchi-identity:
\begin{equation}
\epsilon^{\alpha \beta \mu \nu} \partial_\beta F_{\mu \nu} = 0\,.
\label{Bianchi}
\end{equation}
In $D=3+1$ this identity comprises the well--known 
Maxwell equation $\vec{\partial}\, \cdot\vec{B\,}(\vec{x\,})=0$
stating that magnetic field lines must be closed or extend to
spatial infinity outside the region of interest. The latter scenario 
is not adequate for the study of the vacuum energy which requires to 
integrate over full space. In $D=2+1$ the Bianchi identity becomes
$\frac{\partial B}{\partial t} =  - \partial_x E_y + \partial_y E_x$.
This implies that it is impossible to create (static) net flux 
configurations from zero flux. This may cause inconsistencies as 
we want to compare energies of configurations with and without 
fluxes. We therefore superimpose a \emph{return flux} configuration 
according to
\begin{eqnarray}
B_R(r)&=&-\frac{16 F_G}{\pi R^2\left(1+256\left(
r^2/R^2-1\right)^2\right)\left(\pi/2+\arctan(16)\right)}\cr\cr\cr
B_0 (r)&=&B_G(r) + B_R(r)\,.
\label{returnflux}
\end{eqnarray}
In what follows we will refer to this configuration
as the zero net flux configuration. It is straightforward to verify that
$\lim_{R \rightarrow \infty} E_{\rm cl}[B_0] = E_{\rm cl}[B_G]$
and 
$\lim_{R \rightarrow \infty} E_{\rm FD}[B_0] = E_{\rm FD}[B_G]$.
That is, the return flux does not contribute to the classical
and counterterm energies as the position $R$ of the return flux
is sent to spatial infinity. The crucial question obviously is 
about the behavior of $ E_{\rm vac}[B_0]$ as $R\to\infty$.
To see what happens we compare the integrands of $E_\delta^{(3)}$
for configurations with and without fluxes for two values of $R$ in 
figure~\ref{fig_1}.
The integrand corresponding to $B_0$ oscillates
around that corresponding to $B_G$ and these oscillations 
diminish as $R$ increases. This indicates that indeed
$\lim_{R \rightarrow \infty} E_{\rm vac}[B_0] = E_{\rm vac}[B_G]$.
In order to unambiguously decide on that question we have to 
consider the energy density. This will allow us to distinguish between 
the contributions from the center flux tube and the return flux.

\section{Energy Density}

As motivated above, we consider the vacuum polarization energy 
density 
\begin{equation}
\epsilon(r)=2\pi r\langle T_{00}\rangle
\label{vacengdens}
\end{equation}
to decide whether or not a vacuum polarization energy can 
be attributed to a single flux tube. Here $\langle T_{00}\rangle$ 
denotes a specific matrix element of the energy momentum tensor
$T_{\mu\nu}$ in the background of the zero net flux configuration.
As for the total energy the energy density is computed from three 
entries,
\begin{equation}
\epsilon(r)=\epsilon_\delta(r)+\epsilon_{\rm FD}(r)+\epsilon_{\rm ct}(r)\,,
\label{vacengdens1}
\end{equation}
the contribution form scattering data, $\epsilon_\delta(r)$ with 
appropriate Born terms subtracted to render the momentum integrals
finite, the Feynman diagram contribution, $\epsilon_{\rm FD}(r)$ as 
the Born subtractions must be added back in and the counterterm
contribution from ${\cal L}_{\rm ct}$, eq.~(\ref{lct}). For the 
details on this computation we refer to ref.~\cite{Gr04}, in particular
for the discussion on how the counterterm contribution cancels the 
UV divergences without the need for additional surface counterterms.
Of course, a general consistency condition is that eq.~(\ref{evacren}) is 
obtained from the spatial integral $\int_0^\infty dr \epsilon(r)$.
This requires to consider zero net flux configurations 
because only then the analytic properties of the scattering data are
guaranteed that underly the proof of that identity, {\it cf.\@}
ref.~\cite{Gr02a}.

\begin{figure}[b]
\begin{center}
\includegraphics[width=6.0cm,height=4.5cm]{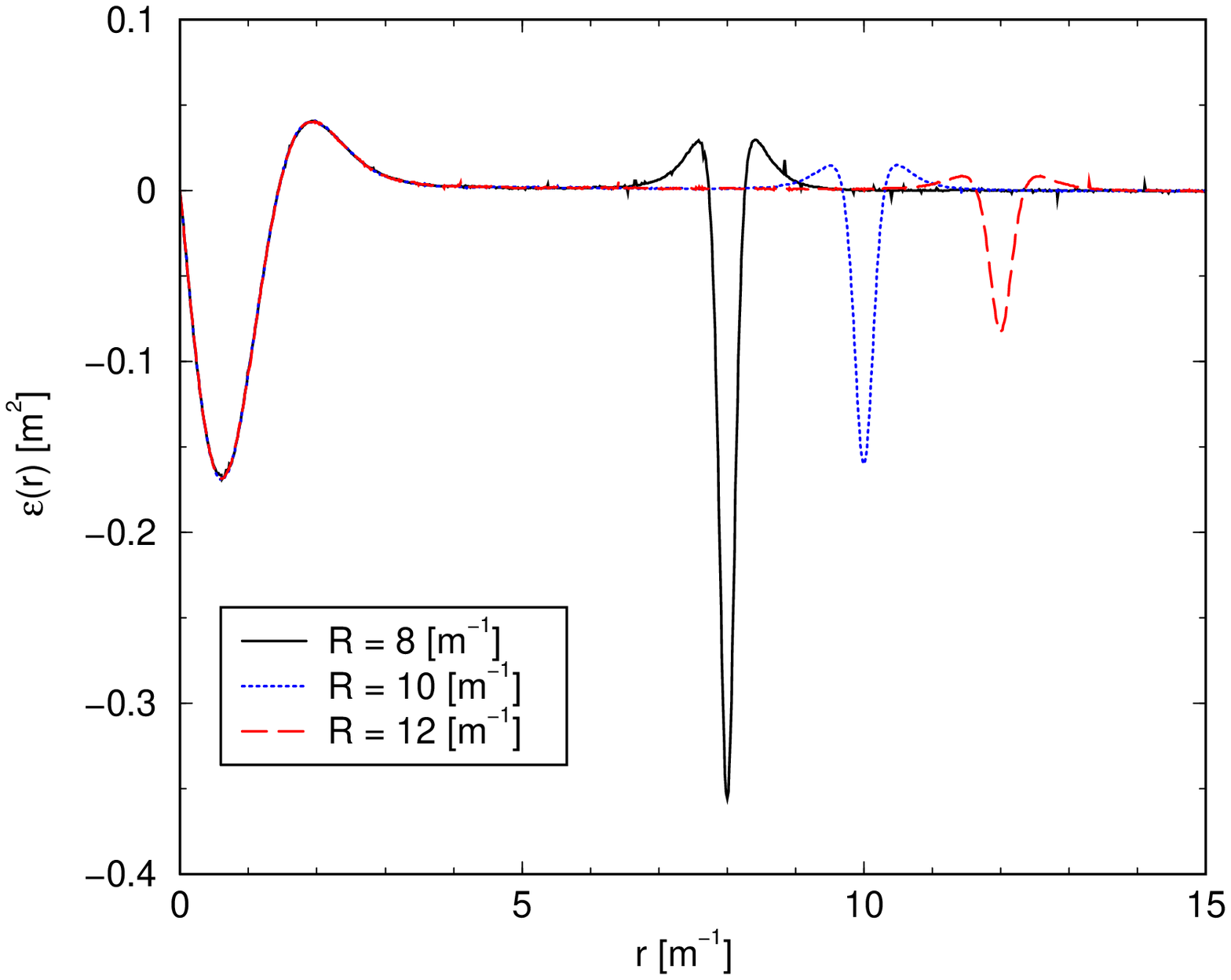}
\hskip1cm
\includegraphics[width=6.0cm,height=4.5cm]{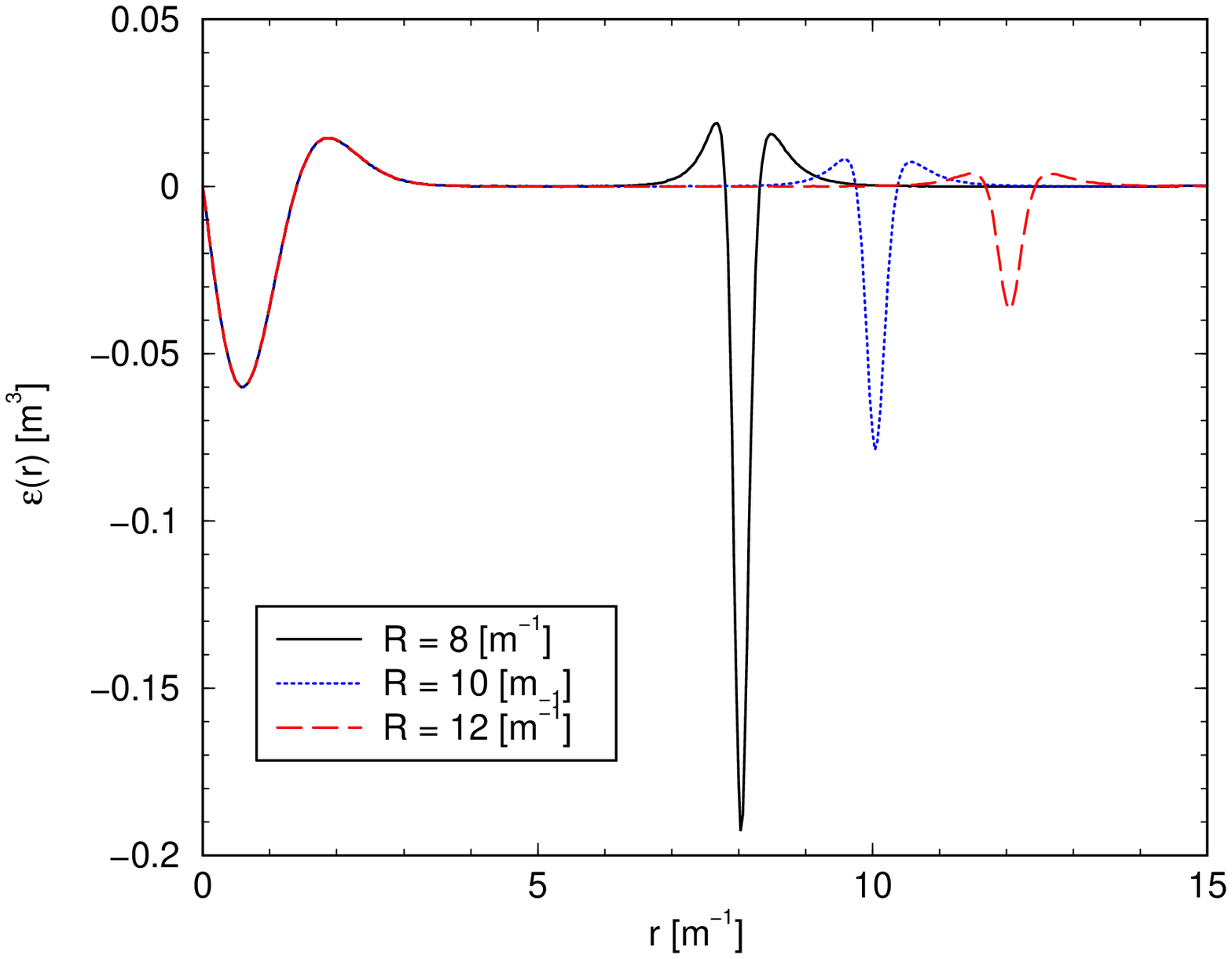}
\caption{\label{fig_2}The left panel shows the  energy 
density $\epsilon(r)$ for various values of the return flux 
position $R$ in $D=2+1$. The right panel is the analog
for the energy density per unit length of the vortex in
$D=3+1$.}
\end{center}
\end{figure}

In figure~\ref{fig_2} we display the energy density as 
a function of the return flux position,~$R$.
A number of conclusions can be drawn from these numerical studies:
1) the energy density in central region of small $r$ is independent of 
$R$, 2) the integrated density from that region gives
$\lim_{R\to\infty}\int^{R/2}_0 dr\, \epsilon(r) = E_{\rm vac}[B_G]$,
3) the energy from the return flux diminishes as its position
is sent to infinity, {\it i.e.\@} 
$\lim_{R\to\infty}\int_{R/2}^\infty dr\,\epsilon(r)=0$.
Altogether this is good news as it clearly confirms the
na{\"\i}ve method that only considers a single central
flux tube. The discontinuities in the phase shifts at $k=0$ do 
not propagate to the vacuum polarization energy.

\section{Approximation Schemes for $E_{\rm vac}[B_G]$}

\begin{figure}[b]
\begin{center}
\includegraphics[width=7.5cm,height=5.5cm]{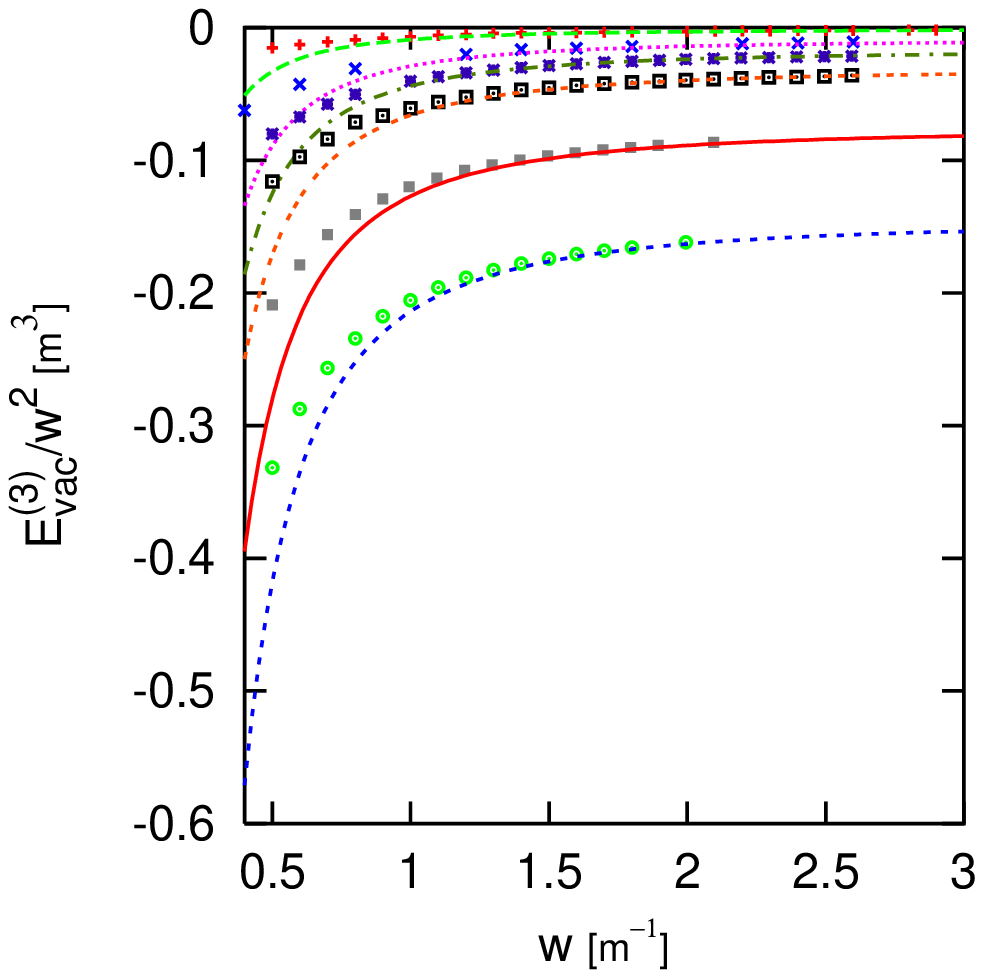}
\hskip 0.2cm
\includegraphics[width=7.5cm,height=5.5cm]{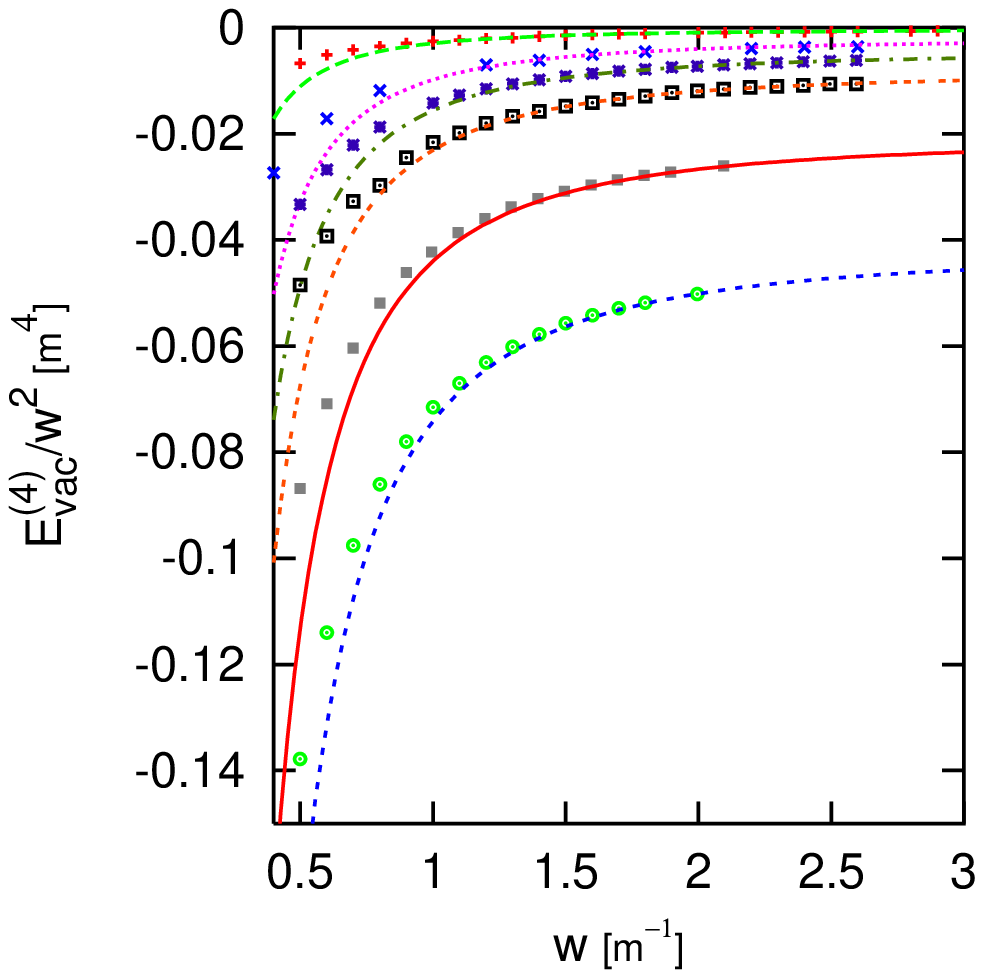}
\caption{\label{fig_3}Renormalized one-loop energies in $D=2+1$
(left panel) and $D=3+1$ (right panel), for fixed values of the magnetic
field at the origin, as a function of the width of the Gau{\ss}ian flux
tube.  The lines correspond to the derivative expansion to second 
order~\cite{Ca95}. The dots, circles, etc. represent the exact results 
for $E_{\rm vac}$. From top to bottom, 
$e B_{\rm G}(0)/m^2 = 1.1, 2, 2.5, 3, 4, 5$.}
\end{center}
\end{figure}

Having established a reliable method for the computation of the 
vacuum polarization energies of flux tubes provides a good
opportunity to employ 
this method to judge approximation schemes. Most popular are
the derivative and perturbative expansion schemes. In both cases 
we will study the dependence of the vacuum polarization energy 
on the width $w$, introduced in eq.~(\ref{gaussft}). 
It is important to consider variations of the background 
field that are consistent with the respective scheme. For 
example, for the derivative expansion to be appropriate
we require configurations that vary slowly but keep the
amplitude $B_G(0)$ fixed. On the contrary, for the 
perturbative expansion we wish to consider different amplitudes
and thus keep the magnetic flux $F_G$, fixed as we change $w$.
The result for the derivative expansion are shown in 
figure~\ref{fig_3}.
In both cases, $D=2+1$ and $D=3+1$, we find good agreement between the 
exact result and the leading contributions in the derivative expansion, 
even though the derivative expansion is known to be an asymptotic 
expansion and hence does not converge when summed 
to all orders~\cite{Dunne}. In the perturbative expansion we
evaluate the leading Feynman diagram,

\smallskip\smallskip

\parbox[t]{12cm}{\hskip-2.8cm
\begin{minipage}[l]{10cm}
\begin{equation}
E_{\rm FD}^{(D)}=
\frac{8 \pi \mathcal{F}^2}{(4 \pi)^{D/2}}
\int_0^\infty dp \left[\int_0^\infty dr \frac{d f(r)}{d r}
J_0(p r)\right]^2 
\int_0^1 dx \frac{x(1-x) p \,
\Gamma(2-D/2)}{[m^2+p^2 x(1-x)]^{2-D/2}}\qquad
\label{FD2}
\end{equation}
\end{minipage}}

\begin{figure}[b]
\begin{center}
\includegraphics[width=6cm,height=4.5cm]{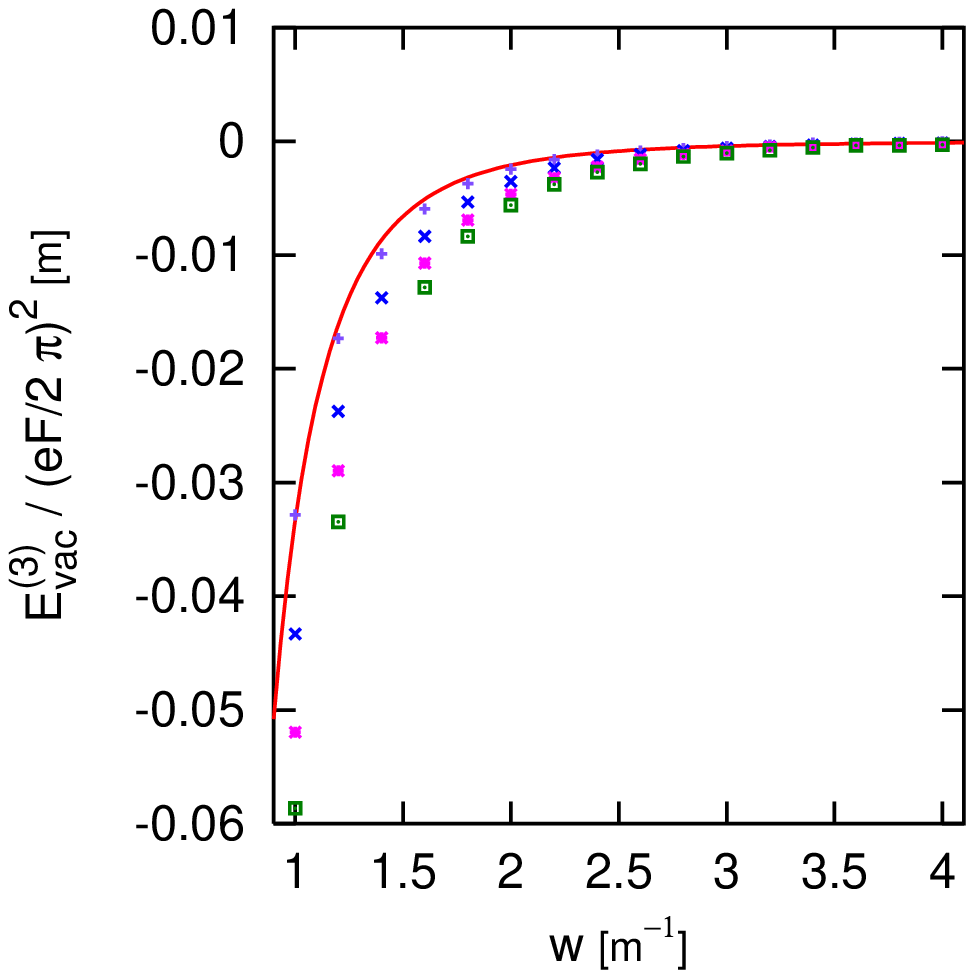}
\hskip 1.0cm
\includegraphics[width=6cm,height=4.5cm]{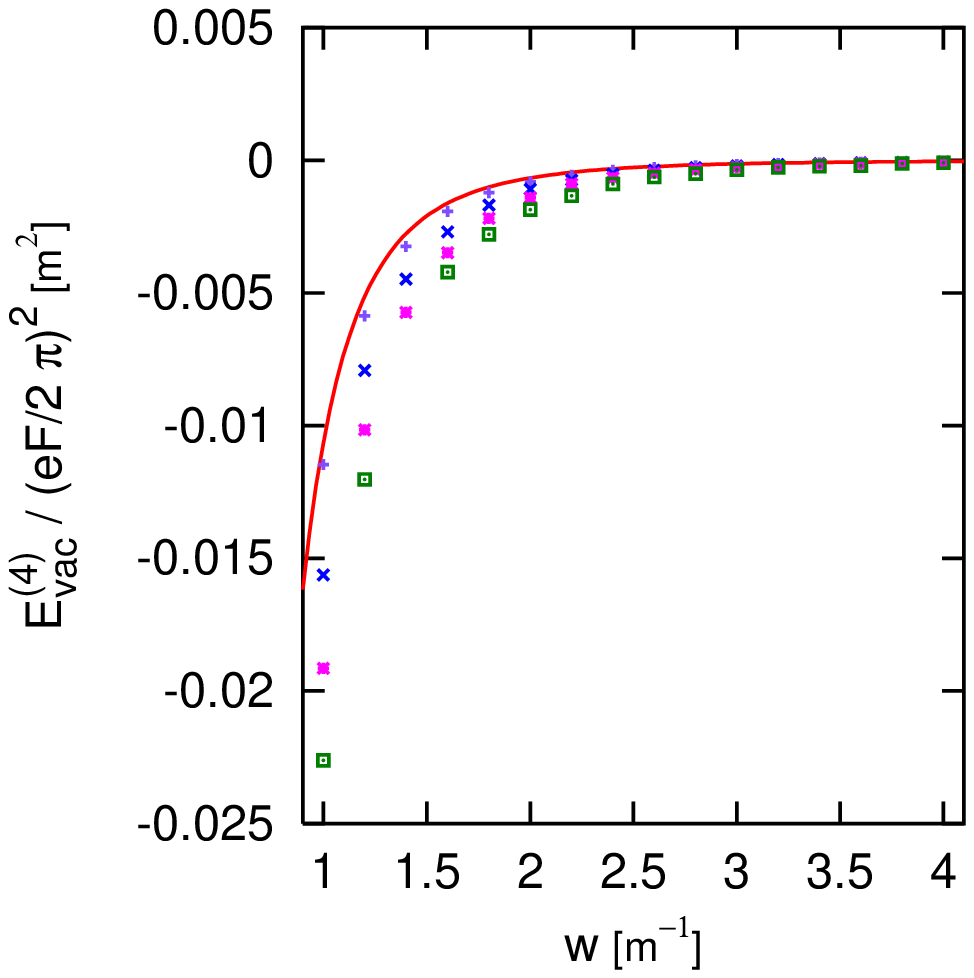}
\caption{\label{fig_4}Renormalized fermion vacuum
polarization energy in units of $\mathcal{F}^2$ as a function of the width,
for various fixed values of the flux $\mathcal{F}$ (2.5, 4.5, 6.5, 8.5
from top to bottom) in the Gau{\ss}ian flux tube. The full line
represents the leading perturbation expansion contribution. The 
left panel is for $D=2+1$ and the right panel for $D=3+1$.}
\end{center}
\end{figure}

\bigskip\noindent
in $D=2+1$ and $D=3+1$ dimensions with the effective expansion
parameter $\mathcal{F}=\frac{e}{2\pi}F_G$. The UV divergence has
not yet been removed from the diagram, eq.~(\ref{FD2}) and the
counterterm part has to be added. This is illuminating especially
for $D=2+1$ because with our renormalization condition it 
changes the leading behavior from $\mathcal{F}/w^2$ to
$\mathcal{F}^2/w^4$. Stated otherwise, renormalization is
essential even for finite quantities. The change in this
leading behavior is crucial to obtain agreement between
the exact results and the perturbative expansion, both in 
$D=2+1$ and $D=3+1$, {\it cf.\@} figure~\ref{fig_4}.

Obviously the approximation schemes work well for the $D=2+1$ and 
$D=3+1$ cases. Furthermore the vacuum polarization energies 
are negative and decrease with the width of the background 
flux tube for both cases. Also the energy densities (per unit length 
in $D=3+1$) are similar in structure as can {\it e.g.\@} be seen from
figure~\ref{fig_2}. As discussed above, to obtain these similarities
it has been crucial to impose \emph{identical} renormalization conditions.
Having done so, it is to be expected that the simpler $D=2+1$ case 
can be utilized to gain information about vacuum polarization energies 
of vortex configurations in more complicated $D=3+1$ problems~\cite{wstring}.

\section{Conclusions}

In this talk I have reported on computations that we have performed 
for one--loop energies and energy densities of electromagnetic flux 
tubes in three and four spacetime dimensions. In general, this vacuum
polarization energy contains ultraviolet divergences and an important 
feature of our approach is that it allows us to impose the standard
renormalization conditions of perturbative quantum electrodynamics.  
Even though the calculation in three spacetime dimensions does not 
suffer from such divergences, a meaningful comparison between three 
and four dimensions can only be made when identical renormalization 
conditions are imposed. The use of 
scattering data to compute the vacuum polarization energy of
an individual flux tube leads to subtleties arising from the
long--range potential associated with the flux tube background,
which does not satisfy the standard conditions of scattering
theory.  Consequently, the scattering data do not necessarily have the
standard analytic properties required to relate the vacuum 
polarization energy to scattering data.
We have therefore considered field 
configurations in which the flux tube is embedded with 
a well-separated return flux so that the total flux vanishes.
We have constructed a limiting procedure in which this return flux
does not contribute to the energy, enabling us to compute the energy
of an isolated flux tube.  

We do not find qualitative differences between 
three and four dimensions for either the energy or energy density,
once identical renormalization conditions have been imposed.  However,
we stress that renormalization in the case of three dimensions proved
essential to this result because the (finite) counterterm
contribution turned out to be large, thus causing sizable
cancellations in the final result.

This study gives an initial step toward understanding flux tubes and
vortices in more complicated  theories. In particular the similarities
between the three and four dimensional cases can be used to determine 
whether quantum corrections stabilize the classically unstable 
strings in the standard model~\cite{wstring}.

\section*{Acknowledgement}

In this talk I have presented results that originate from a 
collaboration with N. Graham, V. Khemani, M. Quandt and
O. Schr\"oder. I highly appreciate their contribution. 
Also support from the DFG under contract We--1254/10--1 is
acknowledged.

\end{document}